\documentclass[prl,twocolumn,preprintnumbers,amsmath,amssymb,showpacs,superscriptaddress,floatfix]{revtex4}

\usepackage{graphicx}
\usepackage{dcolumn}
\usepackage{bm}

\begin{document}


\title{Spectroscopy of stripe order in La$_{1.8}$Sr$_{0.2}$NiO$_{4}$
using resonant soft x-ray diffraction}

\author{C. Sch{\"u}{\ss}ler-Langeheine}%
\affiliation{II. Physikalisches Institut, Universit{\"a}t zu K{\"o}ln,
Z{\"u}lpicher Str.~77, D-50937 K{\"o}ln, Germany}
\author{J. Schlappa}
\affiliation{II. Physikalisches Institut, Universit{\"a}t zu K{\"o}ln,
Z{\"u}lpicher Str.~77, D-50937 K{\"o}ln, Germany}
\author{A.~Tanaka}
 \affiliation{Department of Quantum Matter, ADSM, Hiroshima University,
 Higashi-Hiroshima 739-8530, Japan}
\author{Z. Hu}
\affiliation{II. Physikalisches Institut, Universit{\"a}t zu K{\"o}ln,
Z{\"u}lpicher Str.~77, D-50937 K{\"o}ln, Germany}
\author{C.~F. Chang}
\affiliation{II. Physikalisches Institut, Universit{\"a}t zu K{\"o}ln,
Z{\"u}lpicher Str.~77, D-50937 K{\"o}ln, Germany}
\author{E. Schierle}
\affiliation{Institut f{\"ur} Experimentalphysik, Freie Universit{\"a}t Berlin,
Arnimallee 14, D-14195 Berlin, Germany}
\author{M. Benomar}
\affiliation{II. Physikalisches Institut, Universit{\"a}t zu K{\"o}ln,
Z{\"u}lpicher Str.~77, D-50937 K{\"o}ln, Germany}
\author{H. Ott}
\affiliation{II. Physikalisches Institut, Universit{\"a}t zu K{\"o}ln,
Z{\"u}lpicher Str.~77, D-50937 K{\"o}ln, Germany}
\affiliation{Institut f{\"ur} Experimentalphysik, Freie Universit{\"a}t Berlin,
Arnimallee 14, D-14195 Berlin, Germany}
\author{E. Weschke}
\affiliation{Institut f{\"ur} Experimentalphysik, Freie Universit{\"a}t Berlin,
Arnimallee 14, D-14195 Berlin, Germany}
\author{G. Kaindl}
\affiliation{Institut f{\"ur} Experimentalphysik, Freie Universit{\"a}t Berlin,
Arnimallee 14, D-14195 Berlin, Germany}
\author{O. Friedt}
\affiliation{II. Physikalisches Institut, Universit{\"a}t zu K{\"o}ln,
Z{\"u}lpicher Str.~77, D-50937 K{\"o}ln, Germany}
\affiliation{Laboratoire L\'{e}on Brillouin, CEA-Saclay, 91191 Gif-sur-Yvette Cedex, France}
\author{G.~A. Sawatzky}
\affiliation{Department of Physics and Astronomy, University of British Columbia,
 6224 Agricultural Rd., Vancouver, B.C., V6T 1Z1, Canada}
\author{H.-J.~Lin}
 \affiliation{National Synchrotron Radiation Research Center,
 Hsinchu 30076, Taiwan}
\author{C.~T.~Chen}
 \affiliation{National Synchrotron Radiation Research Center, 
 Hsinchu 30076, Taiwan}
\author{M. Braden}
\affiliation{II. Physikalisches Institut, Universit{\"a}t zu K{\"o}ln,
Z{\"u}lpicher Str.~77, D-50937 K{\"o}ln, Germany}
\author{L. H. Tjeng}
\affiliation{II. Physikalisches Institut, Universit{\"a}t zu K{\"o}ln,
Z{\"u}lpicher Str.~77, D-50937 K{\"o}ln, Germany}%

\date{\today}

\begin{abstract}
Strong resonant enhancements of the charge-order and spin-order
superstructure-diffraction intensities in
La$_{1.8}$Sr$_{0.2}$NiO$_4$ are observed when x-ray energies in
the vicinity of the Ni $L_{2,3}$ absorption edges are used. The
pronounced photon-energy and polarization dependences of these
diffraction intensities allow for a critical determination of the
local symmetry of the ordered spin and charge carriers. We found
that not only the antiferromagnetic order but also the
charge-order superstructure resides within the NiO$_2$ layers;
the holes are mainly located on in-plane oxygens surrounding a
Ni$^{2+}$ site with the spins coupled antiparallel in close
analogy to Zhang-Rice singlets in the cuprates.
\end{abstract}


\maketitle

Doped charge carriers in strongly correlated oxides are often
found to induce ordered periodic arrangements at low
temperatures. The first system where such a phenomenon was
observed by scattering techniques is hole-doped La$_2$NiO$_4$:
Electron and neutron diffraction experiments found in
La$_{2-x}$Sr$_{x}$NiO$_{4+\delta}$ stripe-like superstructures
\cite{chen:93a,tranquada:94a,sachan:95a}, which were interpreted
in terms of charge ordering. The hole-rich stripes form antiphase
domain walls for the antiferromagnetic order on the Ni$^{2+}$
sites and the spacing between them was determined to be inversely
proportional to the hole concentration $n_h = x+2\delta$. Similar
claims have also been made for the isostructural high-$T_c$
superconductors La$_{2-x}$Sr$_x$CuO$_4$ \cite{tranquada:95b},
although recently competing models have been proposed
\cite{hinkov:04a,hanaguri:04a}.

Understanding of the physics behind the formation of the
superstructures in La$_{1.8}$Sr$_{0.2}$NiO$_4$ is still a
difficult issue \cite{hotta:04a}, which is mainly because the
electronic structure related to these stripe phases is not known.
The standard neutron and x-ray scattering techniques used so far
are not directly sensitive to charge: they pick up mainly the
lattice modulations which are induced by doping. It is therefore
not clear whether one should really model the stripe phases as an
ordering of Ni$^{2+}$ and Ni$^{3+}$ ions. Spectroscopic
techniques, for instance, have revealed that doping introduces
substantial amounts of holes in the oxygen band
\cite{kuiper:91a,pellegrin:96a}, thereby raising the question whether
a description in terms of oxygen hole ordering would be more
appropriate. To make things more confusing, there is not even an
agreement to what extent and with which symmetry the oxygen holes
are bound to the Ni ions \cite{pellegrin:96a,sahiner:96a,kuiper:98a}.

In this paper we present results from a new type of spectroscopic
technique, namely resonant diffraction using soft-x-rays. Here
the diffraction involves virtual electronic excitations into
unoccupied intermediate states \cite{luo:93a}, which lead to a
characteristic modulation of the scattering cross section with
photon energy. In particular excitations in the vicinity of the
transition metal $L_{2,3}$ ($2p$$\rightarrow$$3d$) and rare-earth
$M_{4,5}$ ($3d$$\rightarrow$$4f$) absorption edges are known to
be very sensitive to details of the electronic state of the ions
\cite{groot:94a,tholemem:97a,tanaka:94a}. Resonant diffraction is
in fact a combination of spectroscopy and structure
determination, and is therefore the most suitable and natural
technique to study the electronic structure of superstructures in
transition metal and rare-earth systems
\cite{schuessler:01a,abbamonte:02a,wilkins:03a,dhesi:04a,thomas:04a,weschke:04a,abbamonte:04a}.

We carried out resonant diffraction experiments on
La$_{1.8}$Sr$_{0.2}$NiO$_4$ at the Ni $L_{2,3}$ threshold for
both the spin-order and charge-order superstructures. We observed
pronounced photon-energy and polarization dependences of these
diffraction intensities, which allow us to critically determine
the local symmetry of the ordered spin and charge carriers. Using
a quantitative microscopic theoretical model, we find that the
holes are mainly located on the in-plane oxygens surrounding a
Ni$^{2+}$ site with the spins coupled antiparallel, in close
analogy to Zhang-Rice singlets in the cuprates \cite{zhang:88a}.

A single crystal of La$_{1.8}$Sr$_{0.2}$NiO$_{4}$ was grown by a
traveling solvent method and characterized by neutron diffraction
at the Orph\'{e}e reactor, diffractometer 3T.1. The crystal
structure at room temperature is tetragonal with I4/mmm symmetry
and lattice constants $a = 3.807$ {\AA} and $c = 12.5476$ {\AA}.
We will use in the following the larger orthorhombic unit cell
with $a_o \approx b_o = \sqrt{2}a$, which is more commonly used
in the literature. In this notation the diffraction features with
the smallest momentum transfer are the charge-order peak at
(2$\epsilon$,0,1) and the spin-order peak at (1-$\epsilon$,0,0).
For the doping level of our sample, $\epsilon$ is 0.278, and both
peaks can be reached at the Ni $L_{2,3}$ and La $M_{4,5}$
resonances (left panel of Fig. 1).

\begin{figure}[t]
\includegraphics[scale=1,clip,bb=23 419 554 702,width=7cm]{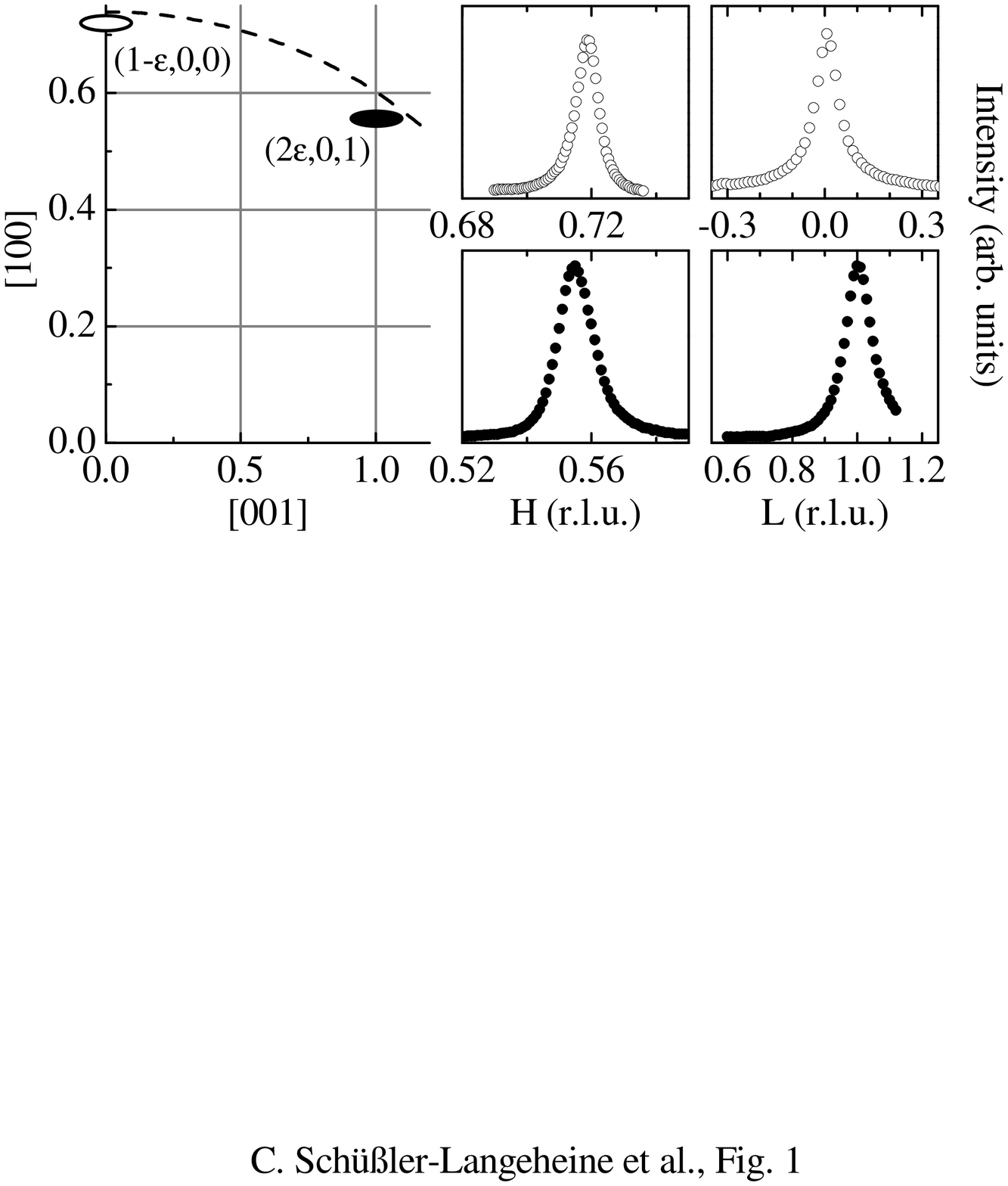}
\caption{Left: Sketch of the superstructure-peak positions in
reciprocal space. The peak at (1-$\epsilon$,0,0) (open symbol)
arises from spin order, the one at (2$\epsilon$,0,1) (filled
symbol) from charge order. The dashed curve denotes the maximum
possible momentum transfer at the Ni $L_3$ resonance. Right:
Scans through the superstructure peaks with $\pi$-polarized light
along the $\left[100\right]$ (H) and $\left[001\right]$ (L)
directions; the upper set shows the spin-order peak, the lower
set the charge-order peak.}
\end{figure}

The soft x-ray scattering experiments were performed at the BESSY
beamlines U49/2-PGM1 and UE52-SGM1, using the two-circle UHV
diffractometer designed at the Freie Universit\"at Berlin \cite{schierle:05a} 
in horizontal scattering geometry. The incoming
light was linearly polarized either parallel to the scattering
plane ($\pi$ polarization) or perpendicular ($\sigma$
polarization) and was monochromatized to an energy resolution of
about 300 meV. A silicon-diode photon detector without
polarization analysis was used with the angular acceptance set to
1 degree in the scattering plane and 5 degrees perpendicular to
it. The sample was cut and polished with a (103) surface
orientation and mounted with the [100] and [001] directions in
the diffraction plane. At 1010 eV photon energy the (002)
structure Bragg peak could be reached and was used together with
the superstructure peaks to orient the sample. Reference
soft-x-ray absorption measurements on
La$_{1.8}$Sr$_{0.2}$NiO$_{4}$ and LaTiO$_3$ have been performed
at the Dragon beamline of the NSRRC in Taiwan.

Fig.~1 shows the position of both superstructure peaks in
reciprocal space, together with scans along the
$\left[100\right]$ (H) and $\left[001\right]$ (L) directions,
taken with the photon energy tuned into the maximum of the Ni
$L_3$ resonance.  From the peak widths we determine the
correlation length in the $ab$ plane for charge order to about
200 {\AA} and for spin order to about 300 {\AA}. The order along
the $c$ direction is less developed with correlation lengths of
50 {\AA} (40 {\AA}) for spin (charge) order.

\begin{figure}[t]
\includegraphics[scale=1,clip,bb=63 204 510 743,width=6.8cm]{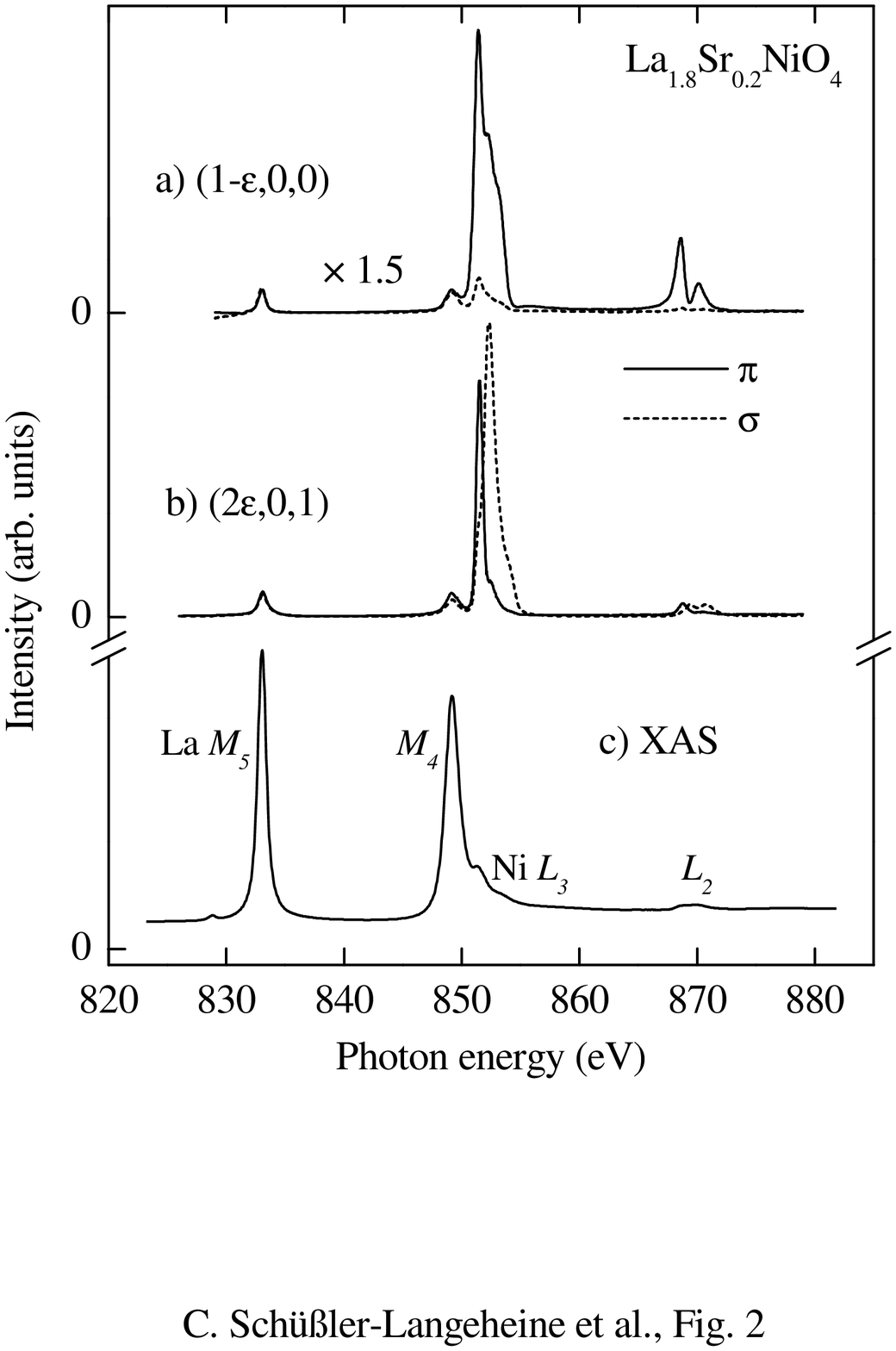}
\caption{Intensities of the a) $(1-\epsilon,0,0)$ spin-order and
b) $(2\epsilon,0,1)$ charge-order superstructure peaks as a
function of photon energy at 60 K for $\pi$- (solid line) and
$\sigma$- (dashed line) polarized light. c) x-ray absorption
signal}
\end{figure}

Fig.~2a) and b) depict the photon-energy dependence of the
superstructure intensities across the La $M_{4,5}$ and Ni
$L_{2,3}$ edges for two different light polarizations
plotted on the same vertical scale \cite{details2}. For both
superstructure peaks the intensity shows a resonant enhancement
mainly for photon energies around 851.6 eV, i.e. in the vicinity
of the Ni $L_3$ white line. This is in striking contrast to the
shape of the XAS data shown in Fig.~2c), which are dominated by
the La $M_5$ resonance at 833 eV and $M_4$ resonance at 849.2 eV.
The gain in contrast for the Ni signal in the diffraction
experiment comes from its sensitivity to only the ordered part of
the system: The strong enhancement at the Ni resonance for both
superstructure signals hence shows directly that not only the
antiferromagnetic order, but also the charge-order superstructure
originates mainly from the NiO$_2$ layers. Both resonances show a
dramatic polarization dependence: For the spin-order peak, the
signal for $\sigma$ polarization is only about 10 percent of that
for $\pi$ polarization; for the charge-order peak, the maxima at
the $L_3$ and $L_2$ resonance when observed with
$\sigma$-polarized light are shifted by about 1 eV towards higher
photon energies as compared to the data taken with $\pi$-polarized
light.

On a closer look, both superstructure peaks show a resonant enhancement at the La
$M_{4,5}$ resonances as well. This is, however, a weak effect
considering the relative intensities of the Ni and La resonances
in the XAS signal. The fact that there is any enhancement of the
charge- and spin-order signal at the La resonance indicates a
coupling between Ni and La sites, which could be caused by the
different bond lengths or by weak hybridization. A
possible strong static dopant order in the La/Sr subsystem can be
ruled out, since the superstructure peaks decrease above 65 K at
both the Ni and La resonances.

The contrast mechanism responsible for the spin-order peak is the
different scattering cross section for different relative
orientations of the electron spins and the polarization
directions of the incoming and outgoing photon. From the weakness
of the signal observed with $\sigma$-polarized light, which in
our scattering geometry means that the electric-field vector of
the incoming photons is parallel to the stripes, one can
conclude, that the Ni spins are essentially (but not perfectly)
collinear with the stripes. The contrast for the direct
observation of charge order arises from the different energy
dependence of the ($2p$$\rightarrow$$3d$) excitation process for
Ni ions with and without the extra holes introduced by doping. The
strong resonance found here demonstrates the extreme sensitivity
of resonant soft-x-ray diffraction.

\begin{figure}[t]
\includegraphics[scale=1,clip,bb=69 135 510 726,width=6.7cm]{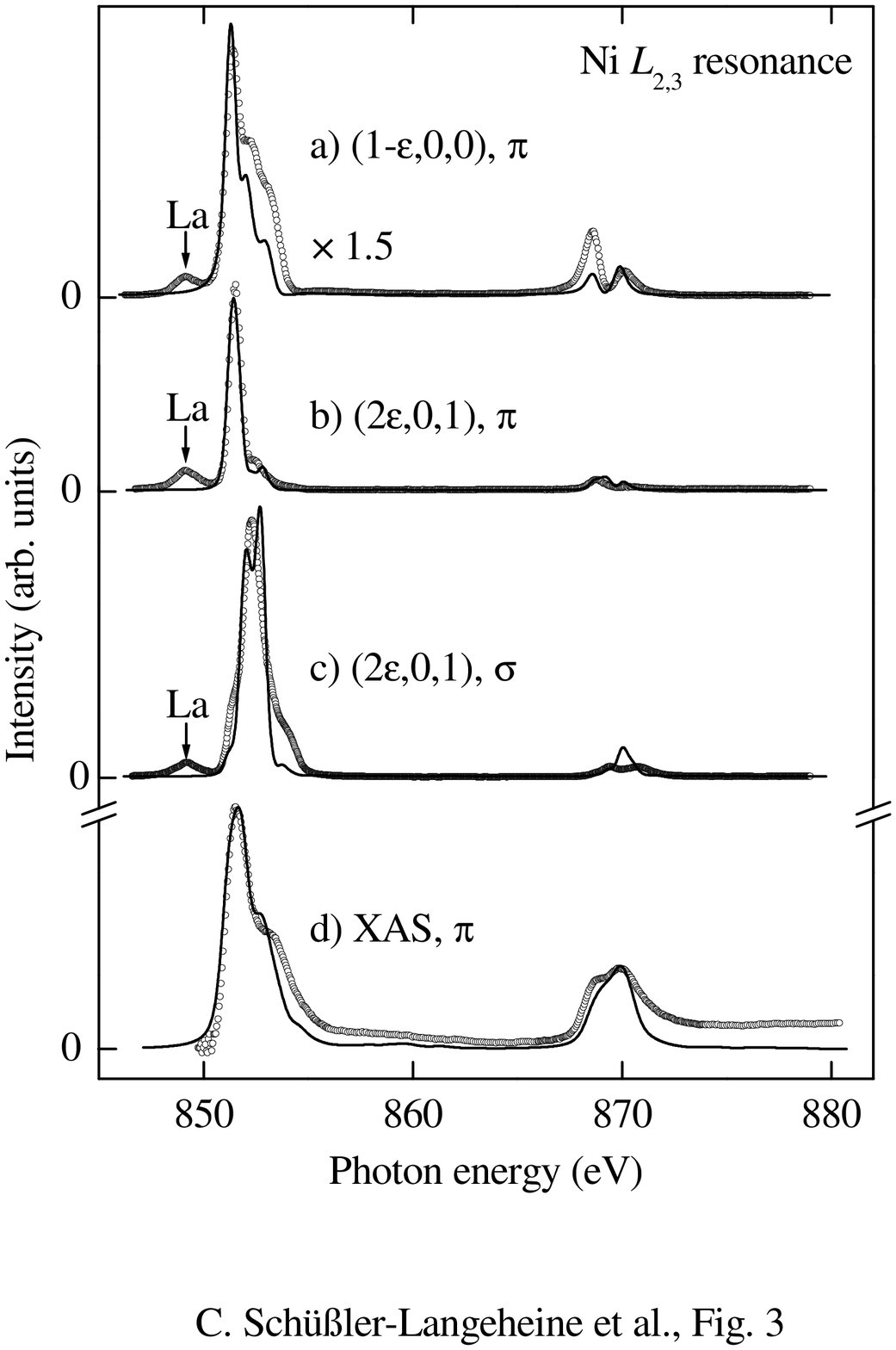}
\caption{Comparison between measured data (symbols) and the
results of the cluster simulation for the spin-order,
$\pi$-polarized light (a); the charge-order superstructure
peak, $\pi$-polarized light (b) and $\sigma$-polarized light (c); 
and the x-ray absorption signal, $\pi$-polarized light (d). 
The La $M_4$ peaks at 849.2 eV, marked by arrows, are not 
included in the Ni $L_{2,3}$ simulation.}
\end{figure}

While some information can be obtained from a qualitative
analysis, e.g., from a comparison of the La and the Ni resonance,
the full power of soft x-ray resonant diffraction evolves from a
combination with a quantitative microscopic theory. We performed
a microscopic modeling based on cluster configuration-interaction
calculations \cite{details1}. The assumed model consists of
diagonal stripes with a periodic order Ni$^{3+}$, Ni$^{2+}$,
Ni$^{2+}$, Ni$^{3+}$, Ni$^{2+}$, Ni$^{2+}$, Ni$^{2+}$, with the
Ni$^{3+}$ stripes serving as antiphase domain walls for the
antiferromagnetic order of the Ni$^{2+}$ spins. Such an
arrangement produces spin- and charge-order diffraction peaks
with $\epsilon =2/7 \approx 0.286$, which is close to the
experimental value. In each Ni$^{2+}$ (Ni$^{3+}$) ion site the
energy-dependent scattering cross section across the resonance
was calculated assuming NiO$_6^{10-}$  (NiO$_6^{9-}$) cluster
with the $D_{4h}$ symmetry, where the ten Ni 3$d$ states and
the nearest-neighbor oxygen 2$p$ states are considered. Coulomb
and exchange interactions between 3$d$ electrons, 3$d$ spin-orbit
interaction, hybridization between the Ni 3$d$ and the oxygen
2$p$ orbitals are taken into account. In the intermediate state
of the scattering process, the Coulomb and exchange interactions
between the 3$d$ electron and 2$p$ core hole and the 2$p$
spin-orbit interaction are necessary to reproduce the
photon energy dependence of the scattering amplitude. For the
simulation of the spin-order signal, only the contributions from
the Ni$^{2+}$ sites were considered and the direction of the
spins was assumed to be parallel to the stripes. The parameters
adopted in the model are $U_{dd}$=7.0 eV, $U_{dc}$=8.5 eV,
$\Delta_{2+}$=6.5 eV (for Ni$^{2+}$), 
$(pd\sigma)$=-1.88 eV (for the oxygens in the
NiO$_2$ plane), $(pd\sigma)$= -1.12 eV (for the apical oxygens)
and $10Dq$=0.5 eV. For the value of $(pd\pi)$, a relation
$(pd\pi)=-\sqrt{3}/4(pd\sigma)$ is assumed.  A Lorentzian
broadening with widths of 0.15 eV and of 0.2 eV at the $L_3$ and
$L_2$ edges (HWHM), respectively, and a Gaussian broadening with
a width of 0.2 eV (HWHM) are applied for the spectra.

With this single set of parameters, we are able to explain the
energy and polarization dependences of the spin- and charge-order
superstructure peaks, including the relative intensities of the 
different diffraction signals (Fig.~3). The model also reproduces 
the experimental Ni contribution to the x-ray
absorption spectrum (Fig.~3d), which was determined by subtracting out the
La contribution obtained from LaTiO$_3$. The agreement
is particularly good for the data obtained with $\pi$ polarization
even in the details of the shape of the resonance; for $\sigma$ 
polarization the agreement in the shape is not that
good, but the model reproduces correctly the 1-eV energy shift
between $\pi$ and $\sigma$ polarization for the charge-order
resonance. Our above mentioned interpretation that the spins and
stripes are parallel aligned, is confirmed by the calculations
since the model gives zero intensity for the spin-order signal
with $\sigma$-polarized light, reproducing the very weak intensity
observed in the experiment.

The model calculations turned out to be very sensitive to the
choice of parameters. The quantitative analysis of the spectra
provides a detailed understanding of the ground state in the
charge-ordered phase: At the Ni$^{3+}$ sites, where the effective 
charge-transfer energy $\Delta_{3+} = \Delta_{2+} - U_{dd}$ is negative, 
the doped holes reside mainly in the ligand molecular orbital with $x^2-y^2$
symmetry. The 3$d$ electron number at the Ni$^{3+}$ ions is as
high as 7.9, which is surprisingly not much smaller than the
value of 8.2 for the 3$d$ electron count on the Ni$^{2+}$ sites.
The spins of the holes on the ligand and 3$d$ orbitals couple
antiferromagnetically and form an $S=1/2$ state
\cite{kuiper:89a,pellegrin:96a} in which the Ni ion with
$t_{2g}^6e_{g\uparrow}^2$ configuration is dressed with a hole in
the $(x^2-y^2)\uparrow$ ligand orbital. This state is analogous
to the Zhang-Rice singlet state in a cuprate superconductor
\cite{zhang:88a}. This assignment is strongly supported by the
observation that between the two polarizations there is a strong
shift of about 1 eV in the peak maxima in each of the $L_2$ and
$L_3$ edges of the $(2\epsilon,0,1)$ spectrum, which indicates a
large energy splitting of the unoccupied $x^2-y^2$ and $3z^2-r^2$
levels. Since the polarization vector of $\pi$-polarized light
has a finite projection on the $c$-axis, the transition
probability of the 2$p$-core electron to the low-lying $3z^2-r^2$
orbital, having lobes along the $c$-axis, is much higher for this
polarization than for $\sigma$ polarization, which has the
polarization vector perpendicular to the $c$-axis. In the
$\sigma$ channel the excitation into the $x^2-y^2$ orbital
dominates. With such a large level splitting, the doped holes can
be considered to be well confined within the in-plane oxygens and
the system possesses strong two-dimensionality like cuprates.

To conclude, we have observed well developed superstructures in
La$_{1.8}$Sr$_{0.2}$NiO$_4$ using resonant soft x-ray
diffraction. The intensities of the spin-order and charge-order
diffraction peaks show strong enhancements when x-ray energies
are tuned to the vicinity of the Ni $L_{2,3}$ absorption edges.
These observations show directly that not only the
antiferromagnetic but also the charge order resides within the
NiO$_2$ layers. The photon-energy and polarization dependence of
the charge and spin-order diffraction intensity across the Ni
$L_{2,3}$ edges can be reproduced by a quantitative microscopic
model calculation assuming diagonal stripes of Ni$^{2+}$ and
Ni$^{3+}$-like ions, in which the Ni$^{3+}$-like objects are formed
of a Ni$^{2+}$ ion accompanied by a hole, which is essentially 
located at the surrounding in-plane
oxygen ions in analogy to a Zhang-Rice singlet in cuprate
systems. The good agreement between experiment and calculation
indicates that differences in the hole concentration around the
Ni ions, i.e. charge ordering, are the cause of the
superstructure.

We gratefully acknowledge the expert support and excellent
working conditions at BESSY. We thank L. Hamdan and R. Bauer for
their skillful technical assistance, T. Koethe for help in preparing the 
experiment, as well as A. Rusydi and P.
Abbamonte for their help during preparational soft-x-ray
diffraction experiments at the National Synchrotron Light Source
at Brookhaven. The research in K\"oln is supported by the
Deutsche Forschungsgemeinschaft through SFB 608, work in Berlin by
the BMBF project 05 KS1 KEE/8.


\begin{thebibliography}{28}
\expandafter\ifx\csname natexlab\endcsname\relax\def\natexlab#1{#1}\fi
\expandafter\ifx\csname bibnamefont\endcsname\relax
  \def\bibnamefont#1{#1}\fi
\expandafter\ifx\csname bibfnamefont\endcsname\relax
  \def\bibfnamefont#1{#1}\fi
\expandafter\ifx\csname citenamefont\endcsname\relax
  \def\citenamefont#1{#1}\fi
\expandafter\ifx\csname url\endcsname\relax
  \def\url#1{\texttt{#1}}\fi
\expandafter\ifx\csname urlprefix\endcsname\relax\def\urlprefix{URL }\fi
\providecommand{\bibinfo}[2]{#2}
\providecommand{\eprint}[2][]{\url{#2}}

\bibitem[{\citenamefont{Chen et~al.}(1993)\citenamefont{Chen, Cheong, and
  Cooper}}]{chen:93a}
\bibinfo{author}{\bibfnamefont{C.~H.} \bibnamefont{Chen}},
  \bibinfo{author}{\bibfnamefont{S.-W.} \bibnamefont{Cheong}},
  \bibnamefont{and} \bibinfo{author}{\bibfnamefont{A.~S.}
  \bibnamefont{Cooper}}, \bibinfo{journal}{Phys.~Rev.~Lett.}
  \textbf{\bibinfo{volume}{71}}, \bibinfo{pages}{2461} (\bibinfo{year}{1993}).

\bibitem[{\citenamefont{Tranquada et~al.}(1994)\citenamefont{Tranquada,
  Buttrey, Sachan, and E.Lorenzo}}]{tranquada:94a}
\bibinfo{author}{\bibfnamefont{J.~M.} \bibnamefont{Tranquada}},
  \bibinfo{author}{\bibfnamefont{D.~J.} \bibnamefont{Buttrey}},
  \bibinfo{author}{\bibfnamefont{V.}~\bibnamefont{Sachan}}, \bibnamefont{and}
  \bibinfo{author}{\bibfnamefont{J.}~\bibnamefont{E.Lorenzo}},
  \bibinfo{journal}{Phys.~Rev.~Lett.} \textbf{\bibinfo{volume}{73}},
  \bibinfo{pages}{1003} (\bibinfo{year}{1994}).

\bibitem[{\citenamefont{Sachan et~al.}(1995)\citenamefont{Sachan, Buttrey,
  Tranquada, Lorenzo, and Shirane}}]{sachan:95a}
\bibinfo{author}{\bibfnamefont{V.}~\bibnamefont{Sachan}},
  \bibinfo{author}{\bibfnamefont{D.~J.} \bibnamefont{Buttrey}},
  \bibinfo{author}{\bibfnamefont{J.~M.} \bibnamefont{Tranquada}},
  \bibinfo{author}{\bibfnamefont{J.~E.} \bibnamefont{Lorenzo}},
  \bibnamefont{and} \bibinfo{author}{\bibfnamefont{G.}~\bibnamefont{Shirane}},
  \bibinfo{journal}{Phys.~Rev.~B} \textbf{\bibinfo{volume}{51}},
  \bibinfo{pages}{12742} (\bibinfo{year}{1995}).

\bibitem[{\citenamefont{Tranquada et~al.}(1995)\citenamefont{Tranquada,
  Sternlieb, Axe, Nakamura, and Uchida}}]{tranquada:95b}
\bibinfo{author}{\bibfnamefont{J.~M.} \bibnamefont{Tranquada}},
  \bibinfo{author}{\bibfnamefont{B.~J.} \bibnamefont{Sternlieb}},
  \bibinfo{author}{\bibfnamefont{J.~D.} \bibnamefont{Axe}},
  \bibinfo{author}{\bibfnamefont{Y.}~\bibnamefont{Nakamura}}, \bibnamefont{and}
  \bibinfo{author}{\bibfnamefont{S.}~\bibnamefont{Uchida}},
  \bibinfo{journal}{Nature (London)} \textbf{\bibinfo{volume}{375}},
  \bibinfo{pages}{561} (\bibinfo{year}{1995}).

\bibitem[{\citenamefont{Hinkov et~al.}(2004)\citenamefont{Hinkov, Pailh\`es,
  Bourges, Sidis, Ivanov, Kulakov, Lin, Chen, Bernhard, and
  Keimer}}]{hinkov:04a}
\bibinfo{author}{\bibfnamefont{V.}~\bibnamefont{Hinkov}},
  \bibinfo{author}{\bibfnamefont{S.}~\bibnamefont{Pailh\`es}},
  \bibinfo{author}{\bibfnamefont{P.}~\bibnamefont{Bourges}},
  \bibinfo{author}{\bibfnamefont{Y.}~\bibnamefont{Sidis}},
  \bibinfo{author}{\bibfnamefont{A.}~\bibnamefont{Ivanov}},
  \bibinfo{author}{\bibfnamefont{A.}~\bibnamefont{Kulakov}},
  \bibinfo{author}{\bibfnamefont{C.~T.} \bibnamefont{Lin}},
  \bibinfo{author}{\bibfnamefont{D.~P.} \bibnamefont{Chen}},
  \bibinfo{author}{\bibfnamefont{C.}~\bibnamefont{Bernhard}}, \bibnamefont{and}
  \bibinfo{author}{\bibfnamefont{B.}~\bibnamefont{Keimer}},
  \bibinfo{journal}{Nature (London)} \textbf{\bibinfo{volume}{430}},
  \bibinfo{pages}{650} (\bibinfo{year}{2004}).

\bibitem[{\citenamefont{Hanaguri et~al.}(2004)\citenamefont{Hanaguri, Lupien,
  Kohsaka, Lee, Azuma, Takano, Takagi, and Davis}}]{hanaguri:04a}
\bibinfo{author}{\bibfnamefont{T.}~\bibnamefont{Hanaguri}},
  \bibinfo{author}{\bibfnamefont{C.}~\bibnamefont{Lupien}},
  \bibinfo{author}{\bibfnamefont{Y.}~\bibnamefont{Kohsaka}},
  \bibinfo{author}{\bibfnamefont{D.-H.} \bibnamefont{Lee}},
  \bibinfo{author}{\bibfnamefont{M.}~\bibnamefont{Azuma}},
  \bibinfo{author}{\bibfnamefont{M.}~\bibnamefont{Takano}},
  \bibinfo{author}{\bibfnamefont{H.}~\bibnamefont{Takagi}}, \bibnamefont{and}
  \bibinfo{author}{\bibfnamefont{J.~C.} \bibnamefont{Davis}},
  \bibinfo{journal}{Nature (London)} \textbf{\bibinfo{volume}{430}},
  \bibinfo{pages}{1001} (\bibinfo{year}{2004}).

\bibitem[{\citenamefont{Hotta and Dagotto}(2004)}]{hotta:04a}
\bibinfo{author}{\bibfnamefont{T.}~\bibnamefont{Hotta}} \bibnamefont{and}
  \bibinfo{author}{\bibfnamefont{E.}~\bibnamefont{Dagotto}},
  \bibinfo{journal}{Phys.~Rev.~Lett.} \textbf{\bibinfo{volume}{92}},
  \bibinfo{pages}{227201} (\bibinfo{year}{2004}).

\bibitem[{\citenamefont{Kuiper et~al.}(1991)\citenamefont{Kuiper, van Elp,
  Sawatzky, Fujimori, Hosoya, and de~Leeuw}}]{kuiper:91a}
\bibinfo{author}{\bibfnamefont{P.}~\bibnamefont{Kuiper}},
  \bibinfo{author}{\bibfnamefont{J.}~\bibnamefont{van Elp}},
  \bibinfo{author}{\bibfnamefont{G.~A.} \bibnamefont{Sawatzky}},
  \bibinfo{author}{\bibfnamefont{A.}~\bibnamefont{Fujimori}},
  \bibinfo{author}{\bibfnamefont{S.}~\bibnamefont{Hosoya}}, \bibnamefont{and}
  \bibinfo{author}{\bibfnamefont{D.~M.} \bibnamefont{de~Leeuw}},
  \bibinfo{journal}{Phys.~Rev.~B} \textbf{\bibinfo{volume}{44}},
  \bibinfo{pages}{4570} (\bibinfo{year}{1991}).

\bibitem[{\citenamefont{Pellegrin et~al.}(1996)\citenamefont{Pellegrin, Zaanen,
  Lin, Meigs, Chen, Ho, Eisaki, and Uchida}}]{pellegrin:96a}
\bibinfo{author}{\bibfnamefont{E.}~\bibnamefont{Pellegrin}},
  \bibinfo{author}{\bibfnamefont{J.}~\bibnamefont{Zaanen}},
  \bibinfo{author}{\bibfnamefont{H.-J.} \bibnamefont{Lin}},
  \bibinfo{author}{\bibfnamefont{G.}~\bibnamefont{Meigs}},
  \bibinfo{author}{\bibfnamefont{C.~T.} \bibnamefont{Chen}},
  \bibinfo{author}{\bibfnamefont{G.~H.} \bibnamefont{Ho}},
  \bibinfo{author}{\bibfnamefont{H.}~\bibnamefont{Eisaki}}, \bibnamefont{and}
  \bibinfo{author}{\bibfnamefont{S.}~\bibnamefont{Uchida}},
  \bibinfo{journal}{Phys.~Rev.~B} \textbf{\bibinfo{volume}{53}},
  \bibinfo{pages}{10667} (\bibinfo{year}{1996}).

\bibitem[{\citenamefont{Sahiner et~al.}(1996)\citenamefont{Sahiner, Croft,
  Zhang, Greenblantt, Perez, Metcalf, Jhans, Liang, and Jeon}}]{sahiner:96a}
\bibinfo{author}{\bibfnamefont{A.}~\bibnamefont{Sahiner}},
  \bibinfo{author}{\bibfnamefont{M.}~\bibnamefont{Croft}},
  \bibinfo{author}{\bibfnamefont{Z.}~\bibnamefont{Zhang}},
  \bibinfo{author}{\bibfnamefont{M.}~\bibnamefont{Greenblatt}},
  \bibinfo{author}{\bibfnamefont{I.}~\bibnamefont{Perez}},
  \bibinfo{author}{\bibfnamefont{P.}~\bibnamefont{Metcalf}},
  \bibinfo{author}{\bibfnamefont{H.}~\bibnamefont{Jhans}},
  \bibinfo{author}{\bibfnamefont{G.}~\bibnamefont{Liang}}, \bibnamefont{and}
  \bibinfo{author}{\bibfnamefont{Y.}~\bibnamefont{Jeon}},
  \bibinfo{journal}{Phys.~Rev.~B} \textbf{\bibinfo{volume}{53}},
  \bibinfo{pages}{9745} (\bibinfo{year}{1996}).

\bibitem[{\citenamefont{Kuiper et~al.}(1998)\citenamefont{Kuiper, van Elp,
  Rice, Buttrey, Lin, and Chen}}]{kuiper:98a}
\bibinfo{author}{\bibfnamefont{P.}~\bibnamefont{Kuiper}},
  \bibinfo{author}{\bibfnamefont{J.}~\bibnamefont{van Elp}},
  \bibinfo{author}{\bibfnamefont{D.~E.} \bibnamefont{Rice}},
  \bibinfo{author}{\bibfnamefont{D.~J.} \bibnamefont{Buttrey}},
  \bibinfo{author}{\bibfnamefont{H.-J.} \bibnamefont{Lin}}, \bibnamefont{and}
  \bibinfo{author}{\bibfnamefont{C.~T.} \bibnamefont{Chen}},
  \bibinfo{journal}{Phys.~Rev.~B} \textbf{\bibinfo{volume}{57}},
  \bibinfo{pages}{1552} (\bibinfo{year}{1998}).

\bibitem[{\citenamefont{Luo et~al.}(1993)\citenamefont{Luo, Trammell, and
  Hannon}}]{luo:93a}
\bibinfo{author}{\bibfnamefont{J.}~\bibnamefont{Luo}},
  \bibinfo{author}{\bibfnamefont{G.~T.} \bibnamefont{Trammell}},
  \bibnamefont{and} \bibinfo{author}{\bibfnamefont{J.~P.}
  \bibnamefont{Hannon}}, \bibinfo{journal}{Phys.~Rev.~Lett.}
  \textbf{\bibinfo{volume}{71}}, \bibinfo{pages}{287} (\bibinfo{year}{1993}).

\bibitem[{\citenamefont{{de Groot}}(1994)}]{groot:94a}
\bibinfo{author}{\bibfnamefont{F.~M.~F.} \bibnamefont{{de Groot}}},
  \bibinfo{journal}{J. Electron Spectrosc. Relat. Phenom.}
  \textbf{\bibinfo{volume}{67}}, \bibinfo{pages}{529} (\bibinfo{year}{1994}).

\bibitem[{\citenamefont{{See review in the Theo Thole Memorial
  Issue}}(1997)}]{tholemem:97a}
\bibinfo{author}{\bibnamefont{{See review in the Theo Thole Memorial Issue}}},
  \bibinfo{journal}{J. Electron Spectrosc. Relat. Phenom.}
  \textbf{\bibinfo{volume}{86}}, \bibinfo{pages}{1} (\bibinfo{year}{1997}).

\bibitem[{\citenamefont{Tanaka and Jo}(1994)}]{tanaka:94a}
\bibinfo{author}{\bibfnamefont{A.}~\bibnamefont{Tanaka}} \bibnamefont{and}
  \bibinfo{author}{\bibfnamefont{T.}~\bibnamefont{Jo}}, \bibinfo{journal}{J.
  Phys. Soc. Jpn.} \textbf{\bibinfo{volume}{63}}, \bibinfo{pages}{2788}
  (\bibinfo{year}{1994}).

\bibitem[{\citenamefont{Sch\"u{\ss}ler-Langeheine
  et~al.}(2001)\citenamefont{Sch\"u{\ss}ler-Langeheine, Weschke, Grigoriev,
  Ott, Meier, Vyalikh, Mazumdar, Sutter, Abernathy, Gr\"ubel
  et~al.}}]{schuessler:01a}
\bibinfo{author}{\bibfnamefont{C.}~\bibnamefont{Sch\"u{\ss}ler-Langeheine}},
  \bibinfo{author}{\bibfnamefont{E.}~\bibnamefont{Weschke}},
  \bibinfo{author}{\bibfnamefont{A.~Y.} \bibnamefont{Grigoriev}},
  \bibinfo{author}{\bibfnamefont{H.}~\bibnamefont{Ott}},
  \bibinfo{author}{\bibfnamefont{R.}~\bibnamefont{Meier}},
  \bibinfo{author}{\bibfnamefont{D.~V.} \bibnamefont{Vyalikh}},
  \bibinfo{author}{\bibfnamefont{C.}~\bibnamefont{Mazumdar}},
  \bibinfo{author}{\bibfnamefont{C.}~\bibnamefont{Sutter}},
  \bibinfo{author}{\bibfnamefont{D.}~\bibnamefont{Abernathy}},
  \bibinfo{author}{\bibfnamefont{G.}~\bibnamefont{Gr\"ubel}},
  \bibnamefont{and} \bibinfo{author}{\bibfnamefont{G.}~\bibnamefont{Kaindl}},
  \bibinfo{journal}{Journ. Electron. Spectrosc Relat.
  Phenom} \textbf{\bibinfo{volume}{114-116}}, \bibinfo{pages}{953}
  (\bibinfo{year}{2001}).

\bibitem[{\citenamefont{Abbamonte et~al.}(2002)\citenamefont{Abbamonte, Venema,
  Rusydi, Sawatzky, Logvenov, and Bozovic}}]{abbamonte:02a}
\bibinfo{author}{\bibfnamefont{P.}~\bibnamefont{Abbamonte}},
  \bibinfo{author}{\bibfnamefont{L.}~\bibnamefont{Venema}},
  \bibinfo{author}{\bibfnamefont{A.}~\bibnamefont{Rusydi}},
  \bibinfo{author}{\bibfnamefont{G.~A.} \bibnamefont{Sawatzky}},
  \bibinfo{author}{\bibfnamefont{G.}~\bibnamefont{Logvenov}}, \bibnamefont{and}
  \bibinfo{author}{\bibfnamefont{I.}~\bibnamefont{Bozovic}},
  \bibinfo{journal}{Science} \textbf{\bibinfo{volume}{297}},
  \bibinfo{pages}{581} (\bibinfo{year}{2002}).

\bibitem[{\citenamefont{Wilkins et~al.}(2003)\citenamefont{Wilkins, Spencer,
  Hatton, Collins, Roper, Prabhakaran, and Boothroyd}}]{wilkins:03a}
\bibinfo{author}{\bibfnamefont{S.~B.} \bibnamefont{Wilkins}},
  \bibinfo{author}{\bibfnamefont{P.~D.} \bibnamefont{Spencer}},
  \bibinfo{author}{\bibfnamefont{P.~D.} \bibnamefont{Hatton}},
  \bibinfo{author}{\bibfnamefont{S.~P.} \bibnamefont{Collins}},
  \bibinfo{author}{\bibfnamefont{M.~D.} \bibnamefont{Roper}},
  \bibinfo{author}{\bibfnamefont{D.}~\bibnamefont{Prabhakaran}},
  \bibnamefont{and} \bibinfo{author}{\bibfnamefont{A.~T.}
  \bibnamefont{Boothroyd}}, \bibinfo{journal}{Phys.~Rev.~Lett.}
  \textbf{\bibinfo{volume}{91}}, \bibinfo{pages}{167205}
  (\bibinfo{year}{2003}).

\bibitem[{\citenamefont{Dhesi et~al.}(2004)\citenamefont{Dhesi, Mirone, De Nadai,
  Ohresser, Bencok, Brookes, Reutler, Revcolevschi, Tagliaferri, Toulemonde
  et~al.}}]{dhesi:04a}
\bibinfo{author}{\bibfnamefont{S.~S.} \bibnamefont{Dhesi}},
  \bibinfo{author}{\bibfnamefont{A.}~\bibnamefont{Mirone}},
  \bibinfo{author}{\bibfnamefont{C.}~\bibnamefont{De Nadai}},
  \bibinfo{author}{\bibfnamefont{P.}~\bibnamefont{Ohresser}},
  \bibinfo{author}{\bibfnamefont{P.}~\bibnamefont{Bencok}},
  \bibinfo{author}{\bibfnamefont{N.~B.} \bibnamefont{Brookes}},
  \bibinfo{author}{\bibfnamefont{P.}~\bibnamefont{Reutler}},
  \bibinfo{author}{\bibfnamefont{A.}~\bibnamefont{Revcolevschi}},
  \bibinfo{author}{\bibfnamefont{A.}~\bibnamefont{Tagliaferri}},
  \bibinfo{author}{\bibfnamefont{O.}~\bibnamefont{Toulemonde}},
  \bibnamefont{and} \bibinfo{author}{\bibfnamefont{G.}~\bibnamefont{van der Laan}},
  \bibinfo{journal}{Phys.~Rev.~Lett.}
  \textbf{\bibinfo{volume}{92}}, \bibinfo{pages}{056403}
  (\bibinfo{year}{2004}).

\bibitem[{\citenamefont{Thomas et~al.}(2004)\citenamefont{Thomas, Hill,
  Grenier, Kim, Abbamonte, Venema, Rusydi, Tomioka, Tokura, McMorrow
  et~al.}}]{thomas:04a}
\bibinfo{author}{\bibfnamefont{K.~J.} \bibnamefont{Thomas}},
  \bibinfo{author}{\bibfnamefont{J.~P.} \bibnamefont{Hill}},
  \bibinfo{author}{\bibfnamefont{S.}~\bibnamefont{Grenier}},
  \bibinfo{author}{\bibfnamefont{Y.-J.} \bibnamefont{Kim}},
  \bibinfo{author}{\bibfnamefont{P.}~\bibnamefont{Abbamonte}},
  \bibinfo{author}{\bibfnamefont{L.}~\bibnamefont{Venema}},
  \bibinfo{author}{\bibfnamefont{A.}~\bibnamefont{Rusydi}},
  \bibinfo{author}{\bibfnamefont{Y.}~\bibnamefont{Tomioka}},
  \bibinfo{author}{\bibfnamefont{Y.}~\bibnamefont{Tokura}},
  \bibinfo{author}{\bibfnamefont{D.~F.} \bibnamefont{McMorrow}},
  \bibinfo{author}{\bibfnamefont{G.} \bibnamefont{Sawatzky}},
  \bibnamefont{and} \bibinfo{author}{\bibfnamefont{M.} \bibnamefont{van Veenendaal}},
  \bibinfo{journal}{Phys.~Rev.~Lett.}
  \textbf{\bibinfo{volume}{92}}, \bibinfo{pages}{237204}
  (\bibinfo{year}{2004}).

\bibitem[{\citenamefont{Abbamonte et~al.}(2004)\citenamefont{Abbamonte,
  Blumberg, Rusydi, Gozar, Evans, Siegrist, Venema, Eisaki, Isaacs, and
  Sawatzky}}]{abbamonte:04a}
\bibinfo{author}{\bibfnamefont{P.}~\bibnamefont{Abbamonte}},
  \bibinfo{author}{\bibfnamefont{G.}~\bibnamefont{Blumberg}},
  \bibinfo{author}{\bibfnamefont{A.}~\bibnamefont{Rusydi}},
  \bibinfo{author}{\bibfnamefont{A.}~\bibnamefont{Gozar}},
  \bibinfo{author}{\bibfnamefont{P.~G.} \bibnamefont{Evans}},
  \bibinfo{author}{\bibfnamefont{T.}~\bibnamefont{Siegrist}},
  \bibinfo{author}{\bibfnamefont{L.}~\bibnamefont{Venema}},
  \bibinfo{author}{\bibfnamefont{H.}~\bibnamefont{Eisaki}},
  \bibinfo{author}{\bibfnamefont{E.~D.} \bibnamefont{Isaacs}},
  \bibnamefont{and} \bibinfo{author}{\bibfnamefont{G.~A.}
  \bibnamefont{Sawatzky}}, \bibinfo{journal}{Nature (London)}
  \textbf{\bibinfo{volume}{431}}, \bibinfo{pages}{1078} 
  (\bibinfo{year}{2004}).

\bibitem[{\citenamefont{Weschke et~al.}(2004)\citenamefont{Weschke, Ott,
  Schierle, Sch\"u{\ss}ler-Langeheine, Vyalikh, Kaindl, Leiner, Ay, Schmitte,
  Zabel et~al.}}]{weschke:04a}
\bibinfo{author}{\bibfnamefont{E.}~\bibnamefont{Weschke}},
  \bibinfo{author}{\bibfnamefont{H.}~\bibnamefont{Ott}},
  \bibinfo{author}{\bibfnamefont{E.}~\bibnamefont{Schierle}},
  \bibinfo{author}{\bibfnamefont{C.}~\bibnamefont{Sch\"u{\ss}ler-Langeheine}},
  \bibinfo{author}{\bibfnamefont{D.~V.} \bibnamefont{Vyalikh}},
  \bibinfo{author}{\bibfnamefont{G.}~\bibnamefont{Kaindl}},
  \bibinfo{author}{\bibfnamefont{V.}~\bibnamefont{Leiner}},
  \bibinfo{author}{\bibfnamefont{M.}~\bibnamefont{Ay}},
  \bibinfo{author}{\bibfnamefont{T.}~\bibnamefont{Schmitte}},
  \bibinfo{author}{\bibfnamefont{H.}~\bibnamefont{Zabel}},
  \bibnamefont{and} \bibinfo{author}{\bibfnamefont{P.~J.}~\bibnamefont{Jensen}},
  \bibinfo{journal}{Phys.~Rev.~Lett.}
  \textbf{\bibinfo{volume}{93}}, \bibinfo{pages}{157204}
  (\bibinfo{year}{2004}).

\bibitem[{\citenamefont{Zhang and Rice}(1988)}]{zhang:88a}
\bibinfo{author}{\bibfnamefont{F.~C.} \bibnamefont{Zhang}} \bibnamefont{and}
  \bibinfo{author}{\bibfnamefont{T.~M.} \bibnamefont{Rice}},
  \bibinfo{journal}{Phys.~Rev.~B} \textbf{\bibinfo{volume}{37}},
  \bibinfo{pages}{R3759} (\bibinfo{year}{1988}).

\bibitem[{sch()}]{schierle:05a}
\bibinfo{howpublished}{E. Weschke, E. Schierle et al., to be published}.

\bibitem[{det({\natexlab{a}})}]{details2}
\bibinfo{howpublished}{The change of the probing volume caused by the variation of
  the photon mean-free path and incidence and detection angle across the resonance 
  has been corrected for using the absorption
  coefficient $\mu$ as obtained from the x-ray absorption spectrum
  \cite{als-nielsen:01a}.}

\bibitem[{det({\natexlab{b}})}]{details1}
\bibinfo{howpublished}{Details of the model calculations will be presented
  elsewhere}.

\bibitem[{\citenamefont{Kuiper et~al.}(1989)\citenamefont{Kuiper, Kruizinga,
  Ghijsen, Sawatzky, and Verweij}}]{kuiper:89a}
\bibinfo{author}{\bibfnamefont{P.}~\bibnamefont{Kuiper}},
  \bibinfo{author}{\bibfnamefont{G.}~\bibnamefont{Kruizinga}},
  \bibinfo{author}{\bibfnamefont{J.}~\bibnamefont{Ghijsen}}, 
  \bibinfo{author}{\bibfnamefont{G.~A.} \bibnamefont{Sawatzky}}, \bibnamefont{and}
  \bibinfo{author}{\bibfnamefont{H.}~\bibnamefont{Verweij}}
  \bibinfo{journal}{Phys.~Rev.~Lett.} \textbf{\bibinfo{volume}{62}},
  \bibinfo{pages}{221} (\bibinfo{year}{1989}).

\bibitem[{\citenamefont{Als-Nielsen and McMorrow}(2001)}]{als-nielsen:01a}
\bibinfo{author}{\bibfnamefont{J.}~\bibnamefont{Als-Nielsen}} \bibnamefont{and}
  \bibinfo{author}{\bibfnamefont{D.}~\bibnamefont{McMorrow}},
  \emph{\bibinfo{title}{Elements of Modern X-ray Physics}}
  (\bibinfo{publisher}{Wiley}, \bibinfo{year}{2001}).

\end{thebibliography}

\end{document}